\documentclass[conference]{IEEEtran}

\newcommand {\be}{\begin{equation}}
\newcommand {\ee}{\end{equation}}
\newcommand {\bey}{\begin{eqnarray}}
\newcommand {\eey}{\end{eqnarray}}
\usepackage{epsfig}
\usepackage{amsfonts}
\usepackage{amsmath}
\usepackage{mathbbol}

\begin{document}

\title{Lower bounds on the communication complexity of two-party
(quantum) processes}

\author{\IEEEauthorblockN{Alberto Montina}
\IEEEauthorblockA{Universit\`a della Svizzera Italiana \\
Via G. Buffi 13 \\ 6900 Lugano, Switzerland \\
Email: montia@usi.ch}
\and
\IEEEauthorblockN{Stefan Wolf}
\IEEEauthorblockA{Universit\`a della Svizzera Italiana \\
Via G. Buffi 13 \\ 6900 Lugano, Switzerland \\
Email: wolfs@usi.ch}
}

\maketitle

\begin{abstract}
The process of state preparation, its transmission and subsequent measurement can
be classically simulated through the communication of some amount of classical
information. Recently, we proved that the minimal communication cost is the 
minimum of a convex functional over a space of suitable probability distributions.
It is now proved that this optimization problem is the dual of a geometric
programming maximization problem, which displays some appealing properties.
First, the number of variables grows linearly with the input size. Second, the 
objective function is linear in the input parameters and the variables. Finally,
the constraints do not depend on the input parameters. These properties imply that, 
once a feasible point is found, the computation of a lower bound on the communication
cost in any two-party 
process is linearly complex. The studied scenario goes beyond quantum processes
and includes the communication complexity scenario introduced by Yao.
We illustrate the method by analytically 
deriving some non-trivial lower bounds. Finally, we conjecture the 
lower bound $n 2^n$ for a noiseless quantum channel with capacity
$n$ qubits. This bound can have an interesting consequence in the context of
the recent quantum-foundational debate on the reality of the quantum state.
\end{abstract}

\IEEEpeerreviewmaketitle

\section{Introduction}

In some distributed computational tasks, the communication of qubits can replace 
a much larger amount of classical communication~\cite{buhrman}. In some cases,
the gap between classical and quantum communication can be even exponential.
What is the ultimate limit to the power of a quantum channel? In a two-party
scenario,
a limit in terms of classical communication is provided by the communication 
complexity of the channel. As defined in Ref.~\cite{montina}, this quantity
is the minimal amount of classical communication
required to simulate the process of preparation of a state, its transmission 
through the channel and its subsequent measurement. In general, the sender
and receiver can have some restriction on the states and measurements that
can be used. In Ref.~\cite{montina}, we proved that the communication
complexity of a quantum channel is the minimum of a convex functional over
a suitable space of probability distributions. 

In this paper, we will show that the original minimization problem is the dual 
of a geometric programming maximization problem with inequality constraints.
As Slater's condition~\cite{boyd}
is satisfied, the duality gap is equal to zero. Thus, the new optimization
problem turns out to be equivalent to the original one. Furthermore,
any feasible point of the constraints provide a lower bound to the
communication complexity. The new reformulation has some interesting
features. First, the number of unknown variables scales linearly in the input 
size. Second, the objective function is linear in the input parameters and the 
variables. Finally, the 
constraints are independent of the input parameters defining the channel.
Thus, if we find the maximum for a particular channel, we can still use
the solution to calculate a lower bound for a different channel, which can
be tight for a slight change of the channel. For example, we could evaluate
the communication complexity for a noiseless quantum channel and, then,
we could wish to find a lower bound for a channel with a small noise.
We will use this reformulation of the original minimization problem
to derive analytically a lower bound for the communication complexity
of a noiseless quantum channel followed by two-outcome projective 
measurements with a rank-$1$ event and its complement. Finally, we
conjecture the lower bound $N\log N$, $N$ being the Hilbert space
dimension.

The considered scenarios are a generalization of the following one. A sender, Alice, 
prepares a quantum state $|\psi\rangle$. For the moment we assume that she can 
choose the state among a finite set whose elements are labeled by an index $a$.
Second, Alice sends the quantum state to another party, Bob, through a quantum 
channel. Then, Bob performs a measurement chosen among a given set whose
elements are labeled by an index $b$. Again, for the moment we assume
that $b$ takes a finite number of values between $1$ and $M$.
Finally, Bob gets an outcome $s$. 
In a more abstract setting, we will consider the overall process as a black
box, which we call C-box, described by a general conditional probability $P(s|a,b)$. 
The C-Box has two inputs $a$ and $b$, which are separately chosen by the
two parties and an outcome $s$, which is obtained by Bob.
This setting goes beyond quantum processes. In particular, it includes the 
communication complexity scenario introduced by 
Yao~\cite{yao}, where $s$ takes two values and $P(s|a,b)$ is deterministic.

A C-box can be simulated classically through a classical channel from Alice
to Bob. We call the minimal communication cost {\it communication
complexity}, denoted by ${\cal C}_{ch}$, of the C-box. Here, we employ
an entropic definition of communication cost
(see Refs.~\cite{montina,montina2} for a detailed definition). 
Similarly, the asymptotic communication complexity,
denoted by ${\cal C}_{ch}^{asym}$, of a C-box is the minimal asymptotic communication
cost in a parallel simulation of many copies of the C-box.
In Ref.~\cite{montina}, we proved that the asymptotic communication complexity ${\cal C}_{ch}^{asym}$
is the minimum of a convex functional over a suitable space, $\cal V$, of probability
distributions. Then, we also proved a tight lower and upper bound for the
communication complexity ${\cal C}_{ch}$ in terms of ${\cal C}_{ch}^{asym}$.
Namely, we have that,
\be
{\cal C}_{ch}^{asym}\le {\cal C}_{ch}\le{\cal C}_{ch}^{asym}+2\log_2({\cal C}_{ch}^{asym}+1)+2\log_2e
.
\ee
Note that a lower bound for the ${\cal C}_{ch}^{asym}$ is also a lower bound
for ${\cal C}_{ch}$. Let us define the set $\cal V$.
\newline
{\bf Definition.} Given a C-box $P(s|a;b)$,
the set ${\cal V}$ contains any conditional probability $\rho({\vec s}|a)$
over the sequence ${\vec s}=\{s_1,\dots,s_M\}$
whose marginal distribution of the $b$-th variable is the distribution $P(s|a,b)$ of the
outcome $s$ given $a$ and $b$.
In other words, the set ${\cal V}$ contains any $\rho({\vec s}|a)$ satisfying the
constraints
\be
\label{constraints}
\begin{array}{c}
\rho({\vec s}|a)\ge0,  \\
\sum_{{\vec s},s_b=s} \rho({\vec s}|a)=P(s|a,b),\; \forall  a, b \text{ and } s,
\end{array}
\ee
where the summation is over every component of the sequence $\vec s$, except the
$b$-th component $s_b$, which is set equal to $s$.
\newline
\newline
Then, we proved that
\be
{\cal C}_{ch}^{asym}=\min_{\rho({\vec s}|a)\in{\cal V}} {\cal C}(a\rightarrow{\vec s}),
\ee
where 
\be
{\cal C}(a\rightarrow{\vec s})\equiv \max_{\rho(a)} I({\bf S};A)
\ee
is the capacity of the channel $\rho({\vec s}|a)$, defined as the maximum of the mutual 
information 
\be
I({\bf S};A)=
\sum_{{\vec s},a}\rho({\vec s}|a)
\rho(a)\log_2\frac{\rho({\vec s}|a)}{\sum_{a'}\rho({\vec s}|a')\rho(a')}
\ee
between the input and the output over the space of input probability distributions
$\rho(a)$~\cite{cover}.
As the mutual information is convex and the maximum over a set of convex functions
is still convex~\cite{boyd}, the asymptotic communication complexity is the
minimum of a convex function over the space~$\cal V$. Since the set $\cal V$ is
also convex, the minimization problem is convex.

As the mutual information is convex in $\rho({\vec s}|a)$ and concave in $\rho(a)$,
we have from the minimax theorem that
${\cal C}_{ch}^{asym}=\max_{\rho(a)} {\cal I}_{\rho(a)}$,
where 
\be\label{F_funct}
{\cal I}_{\rho(a)}\equiv \min_{\rho({\vec s}|a)\in{\cal V}} I({\bf S};A)
\ee
is a functional of the distribution $\rho(a)$.
In some cases, it is trivial to find the distribution $\rho_{max}(a)$ maximizing the
functional $\cal I$.
For example, when there is no restriction on the set of states
and measurements that can be used and the channel is noiseless,
we can infer by symmetry that the distribution $\rho_{max}(a)$ is
uniform. 
Thus, if $\rho_{max}$ is known, the computation of ${\cal C}_{ch}^{asym}$
is reduced to the minimization of the mutual information $I({\cal S};A)$, that is,
${\cal C}_{ch}^{asym}= {\cal I}_{\rho_{max}(a)}$.
More generally, even if $\rho(a)$ does not maximize the functional, we have that
${\cal C}_{ch}^{asym}\ge {\cal I}_{\rho(a)}$.
Thus, the computation of ${\cal I}$ with a non-optimal distribution $\rho(a)$
provides a lower bound on the asymptotic communication complexity. Again, let us recall
that a lower bound for the ${\cal C}_{ch}^{asym}$ is also a lower bound
for ${\cal C}_{ch}$.

\section{Duality}

In the following, we will assume that $\rho(a)$ is given and possibly optimal.
Our task is to show that the computation of~${\cal I}$ is the dual of a
geometric programming maximization problem (See Ref.~\cite{boyd} for a
definition of geometric programming and duality).
Namely, the objective function of the new maximization problem is
\be\label{geom_object_funct}
I=\sum_{s,a,b} P(s|a;b)\rho(a)\lambda(s,a,b),
\ee
which has to be maximized with respect to the variables $\lambda(s,a,b)$ under
the inequality constraints
\be\label{constr}
\sum_a\rho(a)e^{\sum_b\lambda(s_b,a,b)}\le 1,\; \forall {\vec s}=(s_1,\dots,s_M).
\ee
The number of variables is equal to the number of input parameters $P(s|a;b)$,
whereas the number of constraints grows exponentially with the number of
measurements. As the problem is convex and Slater's condition~\cite{boyd} is 
satisfied, strong duality holds and the maximum of $I$ under the constraints
(\ref{constr}) is equal to the minimum of its dual. 
\newline
{\bf Theorem.} Given the maximization problem with objective function~(\ref{geom_object_funct})
and inequality constraints~(\ref{constr}), its dual is the minimization of the
objective function
\be\label{obj_dual}
I_{dual}=\sum_{{\vec s},a}\rho({\vec s}|a)
\rho(a)\log_2\frac{\rho({\vec s}|a)}{\sum_{a'} \rho({\vec s}|a')\rho(a')}
\ee
with respect to the variables $\rho({\vec s}|a)$ under the constraint 
$\rho({\vec s}|a)\in{\cal V}$, that is, under the constraints~(\ref{constraints}).
\newline
{\it Proof.}
It is convenient to introduce a further set of variables, $\alpha({\vec s},a)$, and
the constraint
\be
\alpha({\vec s},a)-\sum_b\lambda(s_b,a,b)=0. 
\ee
Through this equation, we recast Ineqs.~(\ref{constr}) as
\be
1-\sum_a\rho(a)e^{\alpha({\vec s},a)}\ge 0.
\ee
The objective function of the dual problem is the maximum of the Lagrangian
\be
\begin{array}{c}
{\cal L}=I+\sum_{\vec s}\eta({\vec s})\left[1-\sum_a\rho(a)e^{\alpha({\vec s},a)}\right] 
\vspace{2mm}\\
+\sum_{{\vec s},a}\rho({\vec s},a)\left[\alpha({\vec s},a)-\sum_b\lambda(s_b,a,b) \right],
\end{array}
\ee
which is a function of the Lagrange multipliers $\eta({\vec s})$ and $\rho({\vec s},a)$
with the constraint
\be\label{constr_eta}
\eta({\vec s})\ge0.
\ee
By differentiating $\cal L$ with respect to $\lambda(s,a,b)$ and $\alpha({\vec s},a)$,
we get the maximization conditions
\bey\label{cond1}
\sum_{{\vec s},s_b=s}\rho({\vec s},a)=P(s|a,b)\rho(a) \\
\label{cond2}
\eta({\vec s})\rho(a) e^{\alpha({\vec s},a)}=\rho({\vec s},a).
\eey
As the left-hand side of the second equation is positive, we have 
the constraint
\be\label{ineq_rho}
\rho({\vec s},a)\ge0.
\ee
From Eqs.~(\ref{cond1},\ref{cond2}), we have that
\be\label{Lmax}
{\cal L}_{max}=\sum_{\vec s}\eta({\vec s})+
\sum_{{\vec s},a}\rho({\vec s},a)\left(\log_2\frac{\rho({\vec s},a)}{\eta({\vec s})\rho(a)}
-1\right),
\ee
which is the objective function of the dual problem. Now, we can analytically
perform the minimization with respect of $\eta({\vec s})$ under the constraint~(\ref{constr_eta}),
and we get
\be
\eta({\vec s})=\sum_a\rho({\vec s},a)\equiv\rho({\vec s}).
\ee
As $P(s|a,b)$ and $\rho(a)$ are normalized, from this equation and Eq.~(\ref{cond1})
we have that $\sum_{{\vec s}}\eta({\vec s})=\sum_{{\vec s},a}\rho({\vec s},a)=1$.
Let us define the new variable $\rho({\vec s}|a)\equiv\rho({\vec s},a)/\rho(a)$. 
From these equations and Eqs.~(\ref{cond1},\ref{ineq_rho},\ref{Lmax}), we have that the
objective function is the function in Eq.~(\ref{obj_dual})
with the constraints~(\ref{constraints}). % Thus, the dual 
% problem is the original minimization problem in Eq.~(\ref{F_funct}).
$\square$

\subsection{Infinite set of states and measurements}

Until now, we have assumed that Alice and Bob can choose one element in a finite 
set of states and measurements, respectively. The maximization 
problem~(\ref{geom_object_funct},\ref{constr}) can be extended to the case of
infinite sets. In particular, if the sets are uncountable and measurable,
the sums over $a$ and $b$ have to be replaced by integrals. For example,
suppose that Alice can prepare any state and Bob can perform any rank-$1$
projective measurement. Let the dimension of the Hilbert space be $N$. The
space of states is a manifold with dimension $2N-1$ including the physically
irrelevant global phase. The space of measurements is defined as the space
of any orthogonal set of $N$ normalized vectors. Let us denote by 
${\cal M}\equiv(|\phi_1,\dots,|\phi_N\rangle)$ an element in this manifold, where $|\phi_j\rangle$
are the vectors of the orthonormal basis. The function in 
Eq.~(\ref{geom_object_funct}) becomes 
\be\label{obj_continuous}
I=\sum_s\int d{\cal M}\int d\psi P(s|\psi,{\cal M})\lambda(s,\psi,{\cal M})
\ee
in the continuous limit, under the assumption that the integration measure is such that
\be
\int d{\cal M}=\int d\psi=1.
\ee
The second equality implies that $\rho(\psi)=1$, as the distribution is uniform
over the space of quantum states. Let us denote by $S:{\cal M}\rightarrow s$ any 
function mapping a measurement $\cal M$ to a value $s$ in the set of possible outcomes.
The constraints~(\ref{constr}) become
\be
\int d\psi e^{\int d{\cal M} \lambda[S({\cal M}),\psi,{\cal M}]}\le1, \;
\forall\; S .
\ee
This constraint can be recast in the form
\be\label{constr_cont}
\int d\psi e^{\sum_s\int_{\Omega_s} d{\cal M} \lambda(s,\psi,{\cal M})}\le1,\;
\forall\; (\Omega_1,\dots,\Omega_N)\in{\cal P},
\ee
where $(\Omega_1,\dots,\Omega_N)\in P$ is any partition of the measurement manifold
so that $\Omega_i\cap\Omega_j=\emptyset$ if $i\ne j$ and $\cup_i\Omega_i$ is
the whole manifold.

Thus, the optimization problem is the maximization of the objective 
function~(\ref{obj_continuous}) under the constraints~(\ref{constr_cont}).

\section{Application: lower bounds}
The solution of the geometric programming maximization problem introduced in the 
previous section gives the asymptotic communication complexity of a quantum channel. 
Furthermore, any feasible point satisfying the inequality constraints provides a 
lower bound on ${\cal C}_{ch}^{asym}$ and ${\cal C}_{ch}$. As an application of the method, 
let us analytically calculate non-trivial lower bounds in the case of noiseless 
channels and two-outcome measurements with a rank-1 event and its complement. 
In particular, we will consider the cases with $N<5$, for which the calculations 
are simpler. It is possible to find non-trivial lower bounds for arbitrary
dimensions by using the same procedure, but the calculations become harder
as a used differentiability property does not hold for $N\ge5$.
The measurement is specified by a vector $|\phi\rangle$ defining the rank-1 event 
$|\phi\rangle\langle\phi|$ and the complement $\mathbb{1}-|\phi\rangle\langle\phi|$.

The objective function and the constraints take the forms
\be
I=\sum_{s=1}^2\int d{\phi}\int d\psi P(s|\psi,\phi)\lambda(s,\psi,\phi),
\ee
\be
\int d\psi e^{\int_{\Omega} d\phi \lambda(1,\psi,\phi)+
\int_{\Omega^c} d\phi \lambda(2,\psi,\phi)}\le1,\;
\forall\; \Omega,
\ee
where $\Omega$ is a subset of the set of measurements $|\phi\rangle$ and 
$\Omega^c$ is its complement. For a noiseless quantum channel, we have that
\be
P(s|\psi,\phi)=\delta_{s,1}|\langle\psi|\phi\rangle|^2+
\delta_{s,2}(1-|\langle\psi|\phi\rangle|^2).
\ee
The constraints can be written in the form
\be\label{constr_new_form}
\int d\psi e^{\int_{\Omega} d\phi \lambda(\psi,\phi)+
\int d\phi \lambda(2,\psi,\phi)}\le1,\;
\forall\; \Omega,
\ee
where $\lambda(\psi,\phi)\equiv\lambda(1,\psi,\phi)-\lambda(2,\psi,\phi)$.

Every $\lambda(i,\psi,\phi)$ satisfying the constraints induces a lower
bound to the asymptotic communication complexity. A simple form for these
functions is
\be\label{lambda_form}
\lambda(i,\psi,\phi)\equiv \alpha_i |\langle\phi|\psi\rangle|^2+\beta_i.
\ee
The constraints are satisfied for a suitable choice of $\alpha_i$ and $\beta_i$.
This is obviously the case for $\alpha_i=\beta_i=0$. 
Let 
$\alpha\equiv\alpha_1-\alpha_2$ and $\beta\equiv\beta_1-\beta_2$. 
It is simple to show that
\be
\int d\phi |\langle\phi|\psi\rangle|^2=1/N.
\ee
Furthermore,
\be
\int d\phi |\langle\phi|\psi\rangle|^4=\frac{2}{N(N+1)}.
\ee
Using these equations, we have that the objective function takes the form
\be\label{zero_err_funct}
I=\frac{\beta}{N}+\frac{2\alpha}{N(N+1)}+\frac{\alpha_2}{N}+\beta_2
\ee
and the constraints become
\be\label{ineq1}
e^{\frac{\alpha_2}{N}+\beta_2+\beta S_\Omega}\int d\psi e^{\alpha\int_\Omega d\phi 
|\langle\psi|\phi\rangle|^2}\le 1 \;\forall\;\Omega,
\ee
where
\be
S_\Omega\equiv \int_\Omega d\phi .
\ee

Taking $\Omega$ equal to the empty set and to the whole set of vectors, we get 
the inequalities
\be
\frac{\alpha_2}{N}+\beta_2\le0, \;\;
\frac{\alpha}{N}+\beta+\frac{\alpha_2}{N}+\beta_2\le0.
\ee
To have a non-trivial lower bound, the objective function has to be positive,
thus, the above inequalities and the positivity of $I$ give the following
significant region of parameters
\be
\alpha\ge0, \;\;
-\frac{2\alpha}{N+1}\le\beta\le-\frac{\alpha}{N+1}.
\ee
In particular, $\alpha$ must be positive.

Using the Isserlis-Wick theorem~\cite{isserlis}
and the positivity of $\alpha$, it is possible to prove that the left-hand side of
constraint~(\ref{ineq1}) is maximal if $\Omega$ is a suitable cone of vectors. \newline
{\bf Claim.} The left-hand side of the Ineq.~(\ref{ineq1}) is maximal
for a set $\Omega$ such that, for some $|\chi\rangle$ and $\theta\in[0,\pi/2]$,
\be
|\phi\rangle\in\Omega \Longleftrightarrow |\langle\chi|\phi\rangle|^2\ge \cos^2\theta.
\ee
In other words, $\Omega$ is a cone with symmetry axis $|\chi\rangle$ and angular
aperture $2\theta$.
\vspace{2mm}
\newline
Let us denote by $\Omega(\theta)$ a cone with angular aperture $2\theta$.
From this claim, we have that constraints~(\ref{ineq1}) are satisfied
for any $\Omega$ if and only if they are satisfied for $\Omega=\Omega(\theta)$, where
$\theta$ is any element in $[0,\pi/2]$.
Thus, we need to evaluate the integral in the exponent of the constraints only
over any cone $\Omega(\theta)$ of unit vectors. Let us denote by $S(\theta)$
the quantity $S_{\Omega(\theta)}$. It is easy to find that
\be
S(\theta)=\sin^{2N-2}\theta.
\ee
Using equation
\be
\int_{\Omega(\theta)} d\phi |\langle\psi|\phi\rangle|^2=S(\theta)\left(\cos^2\theta
|\langle\psi|\chi\rangle|^2
+\frac{\sin^2\theta}{N}\right)
\ee
(See Ref.~\cite{montina2} for its derivation)
and performing the integral over $\psi$ in the constraints~(\ref{ineq1}),
we obtain the inequalities
\be\begin{array}{c}
{\cal F(\theta,\alpha,\beta)}\equiv -S(\theta) \left[ \beta+
\alpha\left(\frac{\sin^2\theta}{N}+\cos^2\theta\right)\right] \\
-\log\frac{(N-1)!-(N-1)\Gamma(N-1,\alpha S(\theta)\cos^2\theta)}{
\left(\alpha S(\theta)\cos^2\theta\right)^{N-1}}\ge \frac{\alpha_2}{N}+\beta_2,
\;\forall \theta.
\end{array}
\ee
where $\Gamma$ is the incomplete gamma function. 

Since the objective function is linear in the unknown variables, its maximum
is attained when the minimum of ${\cal F}(\theta,\alpha,\beta)$ over $\theta$
is strictly equal to $\frac{\alpha_2}{N}+\beta_2$.
Let $\theta_m(\alpha,\beta)$ be the value of $\theta$ such that ${\cal F}$ is minimum.
We have that
\bey
{\cal F}[\theta_m(\alpha,\beta),\alpha,\beta]=\frac{\alpha_2}{N}+\beta_2\\
\label{deriv}
\left. \frac{d {\cal F}(\theta,\alpha,\beta)}{d \theta}\right|_{\theta=\theta_m(\alpha,\beta)}=0,
\eey
the last equation coming from the fact that $\theta_m$ is a stationary point in $\theta$.
Note that, until this point, the input function $P(s|\psi,\phi)$ is not involved
in the calculations, as it appears only in the objective function.

Using the first equation, we can remove $\beta_2$ and $\alpha_2$ from
the objective function and we get
\be\label{I_ab}
I=\frac{\beta}{N}+\frac{2\alpha}{N(N+1)}+{\cal F}[\theta_m(\alpha,\beta),\alpha,\beta].
\ee
Now, we assume that the function $\theta_m(\alpha,\beta)$ is differentiable in
the maximal point. We have checked a posteriori that this turns out to be true for 
$N<5$, but it is false in higher dimensions, which we will not consider here. Thus, if
the objective function is maximal, then
\be
\frac{\partial I}{\partial\alpha}=0, \;
\frac{\partial I}{\partial\beta}=0.
\ee
With Eq.~(\ref{deriv}), we have three equations and three unknown values,
that is, $\alpha$, $\beta$ and $\theta_m$. To find an analytical solution,
we introduce an approximation by neglecting the gamma function
in ${\cal F(\theta,\alpha,\beta)}$. Then, we will check the validity of
this approximation. The analytical solution is
\be\label{explicit_alpha}
\alpha=\frac{N^2(N+1)}{N-(N+1)N^\frac{1}{1-N}}, 
\ee
\be\label{cond_theta}
\sin^{2N-2}\theta_m=\frac{1}{N}, 
\ee
\be\label{sol_beta}
\beta=\left(\frac{\left(1-N^\frac{1}{1-N}\right)^{-1}-1}{N}-2\right)\frac{\alpha}{N+1}.
\ee

Using these equations, we obtain that the maximum is
\be\label{maxim}
I_{max}=(N-1)\log\frac{N(N+1)\left(N^\frac{1}{1-N}-1\right)e^{-1}}{
\left[(1+N)N^\frac{1}{1-N}-N\right]\Gamma^\frac{1}{N-1}(N)}.
\ee
Thus, in base $2$ of the logarithm, we have the lower bounds
$1.14227$, $1.86776$, and $2.45238$ bits for $N=2,3,4$, respectively.
They are higher than the trivial lower bound of $1$ bit, which is
the classical information that can be communicated through the
channel with subsequent two-outcome measurement. They even beat
the trivial bounds obtained in the case of rank-$1$ measurements, 
$\log_22=1$, $\log_23=1.585$, and $\log_24=2$, although we considered
only simulations of a channel with subsequent two-outcome measurements.
\begin{figure}
\epsfig{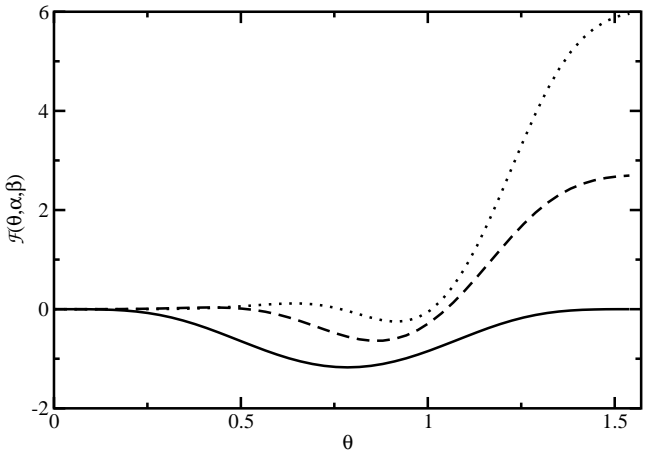}
%\vspace{1mm}
\caption{${\cal F}(\theta,\alpha,\beta)$ as a function of $\theta$ for
$N=2$ (solid line), $N=3$ (dashed line) and $N=4$ (dotted line). The
variables $\alpha$ and $\beta$ take the values maximizing the objective
function $I$ in Eq.~(\ref{I_ab}). The minimum of $\cal F$ is in 
$\theta=\arccos N^\frac{1}{2-2N}$, in agreement with Eq.~(\ref{cond_theta}). }
\label{fig1}
\end{figure}

To derive Eq.~(\ref{maxim}), we have neglected the incomplete gamma function
in ${\cal F}(\theta,\alpha,\beta)$. The exact solution still satisfies
Eqs.~(\ref{cond_theta},\ref{sol_beta}), but the explicit Eq.~(\ref{explicit_alpha})
is replaced by the implicit equation for $\alpha$
\be\label{exact_eq_alpha}
\begin{array}{c}
\left(N^\frac{N}{1-N}-\frac{1}{N+1}\right)\frac{\alpha}{N}+1= \vspace{1mm} 
\frac{e^{-\frac{\cos^2\theta_m\alpha}{N}} \left(\frac{\cos^2\theta_m\alpha}{N}\right)^{N-1}
}{\Gamma(N)-(N-1)\Gamma(N-1,\cos\theta_m\alpha/N)},
\end{array}
\ee
where $\theta_m$ is given by Eq.~(\ref{cond_theta}). The approximate $\alpha$
given by Eq.~(\ref{explicit_alpha}) is obtained by neglecting the right-side
term in Eq.~(\ref{exact_eq_alpha}).

To check the validity of the approximation used to calculate the maximum~(\ref{maxim}),
we have numerically solved the exact Eq.~(\ref{exact_eq_alpha}) through few iterations
of the Newton method. We obtain slightly higher values, thus
Eq.~(\ref{maxim}) gives an exact valid lower bound. The numerical bounds
are $1.14602$, $1.87606$ and $2.46463$ bits for $N=2,3,4$, respectively.
Note that Eq.~(\ref{deriv}) guarantees that $\theta_m$ is a stationary
point of ${\cal F}$, not a minimum. To be sure that $\theta_m$ is actually
a minimum, we have plotted ${\cal F}$ as a function of $\theta$, see
Fig.~\ref{fig1}.

\begin{figure}
\epsfig{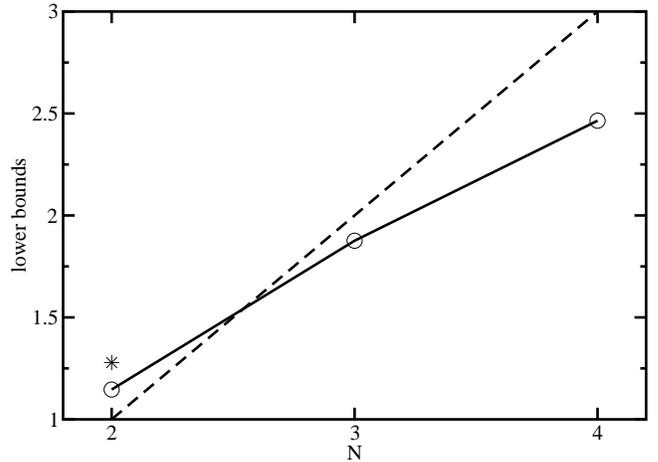}
%\vspace{1mm}
\caption{Calculated lower bound of the communication cost (solid line). The dashed line 
is the lower bound proved using the double cap conjecture~\cite{montina3}. The
star for $N=2$ is the lower bound obtained in Ref.~\cite{montina}. As the measurements
have two outcomes and the quantum channel is noiseless, $1$ bit is a trivial lower bound.}
\label{fig2}
\end{figure}

The lower bound for $N=2$ is lower than the bound $1+\log_2\frac{\pi}{e}\simeq1.2088$
previously derived by us in Ref.~\cite{montina}. A better result can be obtained
by using a slightly different form of $\lambda(i,\psi,\phi)$ in Eq.~(\ref{lambda_form}).
This will be discussed in a more detailed paper~\cite{montina4}.
Also, the other two bounds are lower than the bound $N-1$ proved by one of us 
in Ref.~\cite{montina3}, but the proof
relies on an unproved property, called double-cap conjecture. The overall
bounds are plotted in Fig.~\ref{fig2}. If we extrapolated Eq.~(\ref{maxim}),
we would have that the lower bound for high $N$ would scale as
\be
I_{max}\sim N \log\left(1+\frac{1}{\log N}\right)\sim N/\log N,
\ee
which is sublinear in $N$. Although this asymptotic behavior is not reliable, as
Eq.~(\ref{maxim}) does not hold for $N>4$, it is likely that 
the maximization of $I$ in Eq.~(\ref{I_ab}) for $N>4$ will give a 
result close to Eq.~(\ref{maxim}). 
The case $N>4$ will be discussed in the detailed paper~\cite{montina4}. 
There are some reasons, related to Eq.~(\ref{cond_theta}), suggesting
that the stronger lower bound $N\log N$ for the communication cost can
be achieved with a suitable choice of $\lambda(i,\alpha,\beta)$.

\section{Conclusion}

We have shown that the (asymptotic) communication complexity of a quantum
channel is the maximum of a linear objective function under inequality
constraints. Feasible points of the constraints provide lower bounds on
the communication cost. We have used this optimization problem to derive
analytically some non-trivial lower bounds for a noiseless quantum channel
and subsequent two-outcome measurements with a rank-$1$ event and its
complement. We explicitly evaluated the bounds for a Hilbert space
dimension $N$ between $2$ and $4$. In a more detailed paper~\cite{montina4},
we will discuss the case $N>4$. There are some reasons suggesting
that it is possible to prove the lower bound $N\log N$ with a suitable
choice of $\lambda(i,\psi,\phi)$. This lower bound would have interesting
consequences in the context of the recent debate on the reality of the quantum
state~\cite{pusey,montina5}. The relation between this quantum foundational problem and 
communication complexity was pointed out in Ref.~\cite{montina5}.

{\it Acknowledgments.} 
This work is supported by the Swiss National Science Foundation, 
the NCCR QSIT, and the COST action on Fundamental Problems in Quantum Physics.

\bibliography{biblio.bib}

\end{document}